\documentclass[reprint,aps,superscriptaddress]{revtex4-1}

\usepackage[pdftex]{graphicx}
\usepackage{amsmath}
\usepackage{amsthm}
\usepackage{amssymb}
\usepackage{color}

\newcommand{\ket}[1]{|#1\rangle}
\newcommand{\bra}[1]{\langle#1|}
\newcommand{\ketbra}[2]{|#1\rangle\langle#2|}
\newcommand{\braket}[2]{\langle #1|#2\rangle}
\newcommand{\proj}[1]{\ketbra{#1}{#1}}

\newcommand{\eqrefF}{(19)}

\def\id{\mathsf{id}}
\def\tr{\textrm{tr}}
\def\mms{\pi}

\def\cp{{\cal P}}
\def\cc{{\cal C}}
\def\cent{{{\cal C}_e}}
\def\ch{{{\cal C}_H}}
\def\cq{{\cal Q}}

\def\deph{\bar D}
\def\M{{\cal M}}
\def\N{{\cal N}}

\def\E{{\cal E}}

\def\T{{\cal T}}
\def\G{{\cal G}}

\def\ic{\cq^{(1)}}

\begin{document}
\title{Non-convexity of private capacity and classical environment-assisted capacity of a quantum channel}

\newcommand{\DAMTP}{Department of Applied Mathematics and Theoretical Physics, University of Cambridge, Cambridge CB3 0WA, U.K.}

\author{David Elkouss}
\affiliation{QuTech, Delft University of Technology, Lorentzweg 1, 2628 CJ Delft, The Netherlands}
\author{Sergii Strelchuk}
\affiliation{\DAMTP}

\begin{abstract}
The capacity of classical channels is convex. This is not the case for the quantum capacity of a channel: the capacity of a mixture of different quantum channels exceeds the mixture of the individual capacities and thus is non-convex. Here we show that this effect goes beyond the quantum capacity and holds for the private and classical environment-assisted capacities of quantum channels. 
\end{abstract}
\maketitle


\section{Introduction}

Classical information theory was laid down by Shannon in the nineteen forties to characterize the ultimate rate at which one could hope to transmit classical information over a classical communication channel: the channel capacity. Surprisingly in retrospective, not only it achieved its purpose but the capacity of classical channels turned out to comply with all the properties that one could expect for such a quantity: it can be efficiently computed~\cite{Arimoto_72,Blahut_72} and it gauges the usefulness of the channel in the presence of any additional contextual channel. It is a natural consequence of additivity and convexity of the capacity in the set of channels. 

With quantum channels complemented by various auxiliary resources, a whole new range of communication tasks  became feasible. Notably, they allow for the transmission of quantum and private classical communication -- tasks beyond the reach of classical channels. 
For most of these tasks, the tools used to prove the capacity theorems in the classical case can be generalized.  
However, computability, additivity, and convexity --- the three convenient properties of the classical capacity of classical channels --- do not necessarily translate to the quantum case. In Table~\ref{tableofcaps} we summarize what is known about these properties for a set of relevant quantum channel capacities.

With the exception of the entanglement-assisted capacity \cite{Adami_97, Bennett_02}, there is no known algorithm to compute any of these capacities. It is due to their characterization which in most cases is given by a regularized formula \cite{Lloyd_97,Schumacher_97,Holevo_98,Shor_02,Cai_04,Devetak_05,Medeiros_05, guha_quantum_2014, Karumanchi_14,karumanchi_classical_2016,Winter_16}. Moreover, even non-regularized quantities are notoriously hard to compute. For instance, the Holevo information is known to be NP-complete~\cite{Beigi_07}. 

A capacity is non-additive as a function of a channel if for a given pair of channels the sum of their individual capacities is strictly smaller than the capacity of another channel which is constructed by using both channels in parallel. Hence, a non-additive capacity is contextual: the usefulness of a channel for communication depends on what other channels are available. The private and quantum capacities are known to be non-additive~\cite{Smith_08,Li_09,Smith_09b}. This observation motivated authors in~\cite{Winter_16} to define a new quantity -- the potential capacity -- which characterizes the usefulness of a channel used in parallel with the best possible contextual channel. 

\begin{table}
\begin{center}
\begin{tabular}{ c | c | c | c | c}
     & \textrm{Computability} & \textrm{Additivity} & \textrm{Convexity} \\
     \hline
    $\cq$ \, &? & No~\cite{Smith_08} &  No~\cite{Smith_08} \\
    $\cp$ \, & ?  & No \cite{Li_09,Smith_09b} & {\underline{No}} \\ 
    $\cc$ \, & ? & ? &? \\
    $\cent$ \, & Yes \cite{Bennett_02}  & Yes \cite{Adami_97} & Yes \cite{Adami_97, Bennett_02}\\  
    $\ch$ \, & ?  &No \cite{karumanchi_classical_2016}   & {\underline{No}} 
\end{tabular}
\caption{Main properties of quantum channel capacities: convexity, additivity, and computability. We consider quantum capacity $\cal Q$, private capacity $\cal P$ and unassisted, entanglement-assisted and environment-assisted classical capacities, $\cc$, $\cent$ and $\ch$ respectively.}\label{tableofcaps}
\end{center}
\end{table}

Another important property of the capacities of quantum channels is convexity. The capacity $\T$ of a quantum channel $\N$ is non-convex if there exists a pair of channels $\N_1$ and $\N_2$ and $p\in(0,1)$ such that: 
\begin{equation}
\label{eq:nonconvexitydef}
p\T\left(\N_1\right)+(1-p)\T\left(\N_2\right)<\T\left(p\N_1+(1-p)\N_2\right).
\end{equation} 
In a same vein, non-convexity also implies that capacity is contextual. For a channel $\N$, a contextual channel $\M$ and a mixing parameter $p\in(0,1)$ we can define a non-convexity functional:
\begin{equation}
\label{eq:functional}
\G_{p,\M}(\N) = 1/p\left[\T(p\N+(1-p)\M) - (1-p)\T(\M)\right]
\end{equation}
analogous to the one defined in \cite{Winter_16} for non-additivity. 
This functional induces a (new) potential capacity given by the maximization of $\G_{p,\M}(\N)$ over all contextual channels $\M$ and $p\in(0,1]$. If $\T$ is non-convex, then there exists a channel $\N$ such that its potential capacity is strictly larger than $\T(\N)$ or, equivalently, there exists a triple $p, \N, \M$ for which $\G_{p,\M}(\N)> \T(\N)$.

\begin{figure}
\begin{center}\includegraphics[trim= 0 1.5cm 2cm 0,width=8cm]{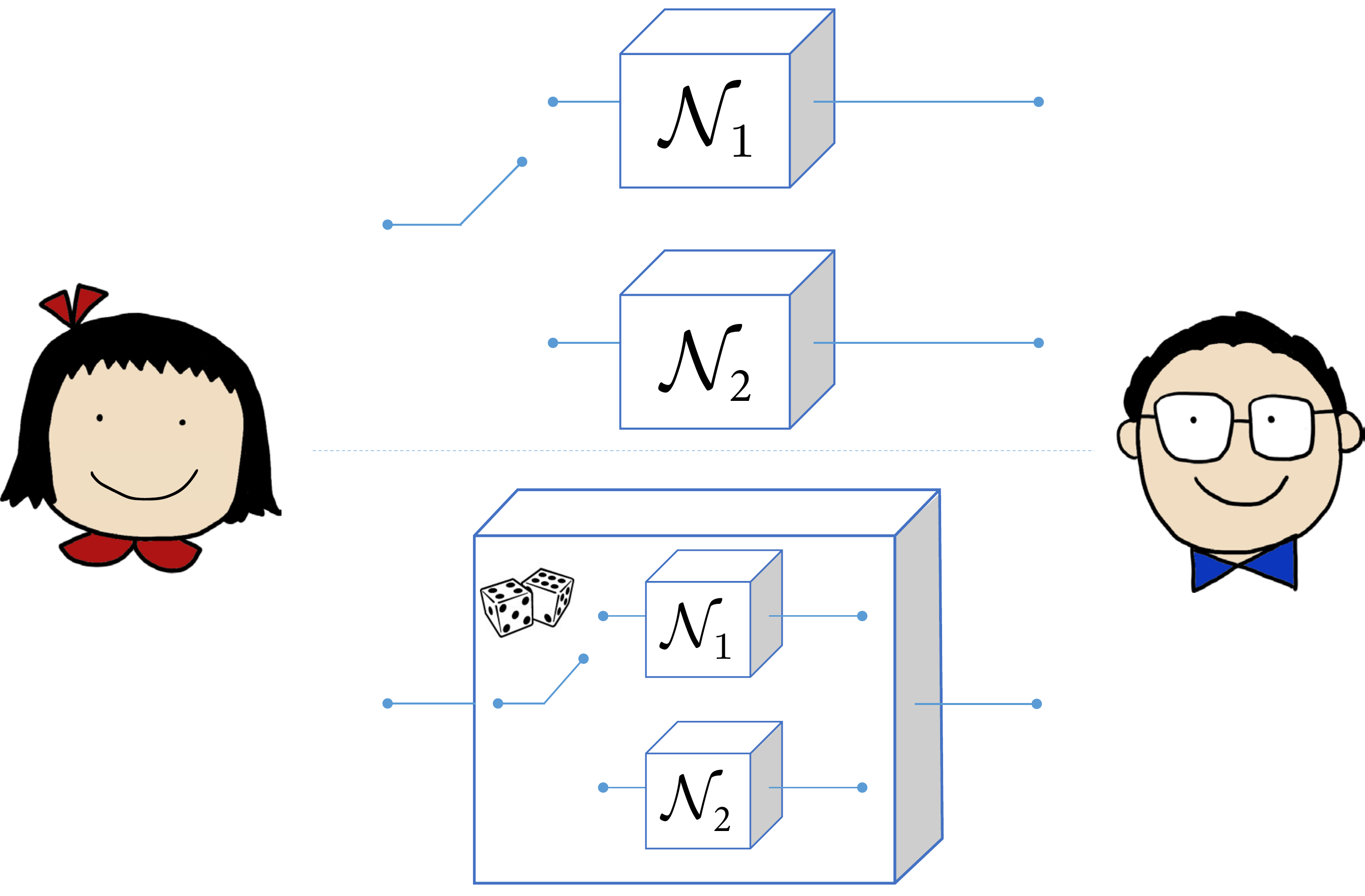}\end{center}
\caption{Operational interpretation of non-convexity. Above, Alice has full control over which channel is applied in the transmission, but she has to apply each channel with some probability. Below, a black box chooses the channel for Alice (with the same probabilities). Non-convexity implies that Alice might communicate at a strictly higher rate in the scenario below.}
\label{fig:prot}
\end{figure}

Non-convexity is a surprising property in connection to two communication scenarios in which Alice, the sender, has access to two channels  
that are used with probabilities $p$ and $1-p$. In the first one, Alice uses both channels independently. In the second one, Alice encodes jointly over the two channels but has no control over which of the channels is applied; instead, a black box applies them at random with the same probabilities $p$ and $1-p$.  
The two scenarios are depicted in Fig.~\ref{fig:prot}.

In contrast to the classical capacity of classical channels, it was shown that the capacity of quantum channels for transmitting quantum information, i.e. the quantum capacity, is non-convex \cite{Smith_08}. The question that we address in the following is whether non-convexity is limited to the transmission of quantum information or can be observed beyond the task of entanglement transmission. We show that the private capacity and the classical environment-assisted capacities of a quantum channel are non-convex.

\section{Communication tasks}

The action of a quantum channel can always be defined by an isometry $V$ that takes the input system $A'$ to the output $B$ together with an auxiliary system called the environment $E$: $\N^{A'\rightarrow B}(\rho^{A'}) =\tr_E V^{A'\rightarrow BE}\rho^{A'} (V^{A'\rightarrow BE})^\dagger$. This isometry allows to define the action of the complementary channel: $\hat\N^{A'\rightarrow E}(\rho^{A'}) =\tr_B V^{A'\rightarrow BE}\rho^{A'} (V^{A'\rightarrow BE})^\dagger$. 
We denote the systems involved by a superscript, which we omit when they are clear from the context.  

Let $\rho^A$ be a quantum state, we denote by $H(A)=-\tr \rho\log\rho$ the von Neumann entropy. Let $\rho^{AB}$ be a bipartite quantum state, we denote by $I(A;B)=H(A)+H(B)-H(AB)$ the mutual information between the systems $A$ and $B$.

We are interested in the following communication tasks and the associated channel capacities. 

The first task is the transmission of quantum information. The quantum capacity characterizes the ability of a quantum channel for this task in the absence of additional resources~\cite{Lloyd_97,Shor_02,Devetak_05}
\begin{equation}
\cq(\N)=\lim_{n\rightarrow\infty}\frac{1}{n}\cq^{(1)}(\N^{\otimes n}),
\end{equation}
where $\cq^{(1)}(\N) = \max_{\phi^{AA'}}\cq^{(1)}(\N,\phi^{AA'})$ is the coherent information of a quantum channel. The maximum is taken over all input states purified with a reference system $A$. The quantity $\cq^{(1)}(\N,\phi^{AA'})=H(B)-H(AB)$, where $H(B),H(AB)$ are the von Neumann entropies of $\rho^B=\N(\tr_{A'}\phi^{AA'})$, $\rho^{AB}=\id^{A}\otimes\N^{A'\rightarrow B}(\phi^{AA'})$ and $\id$ denotes the identity channel. 

For some channels, the coherent information is additive and thus exactly characterizes their capacity. In these cases, it is possible to compute the capacity exactly~\cite{Wilde_13}. However, there are examples when this is not the case~\cite{DiVincenzo_98,Cubitt_15}: coherent information is superadditive. Not only the coherent information is superadditive, but also, the quantum capacity itself is superadditive~\cite{Smith_08,Brandao_12} -- there exist pairs of channels such that their joint capacity is strictly larger than the sum of their capacities.

The second task is the transmission of private classical information. The capacity of a channel for this task without additional resources is called the private capacity~\cite{Cai_04,Devetak_05}. 
We define the private information to be
\begin{equation}
\cp^{(1)}(\N)=\max_{\sum_xp_x\ketbra{x}{x}^X\otimes\rho^{A'}}I(X;B)-I(X;E),
\end{equation}
where $I(X;B)$ and $I(X;E)$ are evaluated on the states $\id^X\otimes\N^{A'\rightarrow B}(\sum_xp_x\ketbra{x}{x}^X\otimes\rho^{A'})$ and $\id\otimes\hat\N^{A'\rightarrow E}(\sum_xp_x\ketbra{x}{x}^X\otimes\rho^{A'})$.
The private capacity is given by the regularization of the private information
\begin{equation}
\cp(\N)=\lim_{n\rightarrow\infty}\frac{1}{n}\cp^{(1)}(\N^{\otimes n}).
\end{equation}
Both private information~\cite{Smith_08b,Kern_08,Elkouss_15} and the private capacity~\cite{Smith_09,Li_09,Smith_09b} were found to be superadditive.

The third task is the transmission of classical information. The classical capacity~\cite{Holevo_98,Schumacher_97} characterizes the capacity of a quantum channel for transmitting classical information without additional resources. To characterize the classical capacity we first define the Holevo information
\begin{equation}
\cc^{(1)}(\N)=\max_{\sum_xp_x\ketbra{x}{x}^X\otimes\rho^{A'}}I(X;B) .
\end{equation}
The classical capacity is given by the regularization of the Holevo information
\begin{equation}
\cc(\N)=\lim_{n\rightarrow\infty}\frac{1}{n}\cc^{(1)}(\N^{\otimes n}).
\end{equation}
Holevo information is superadditive~\cite{Hastings_09} but it is a challenging open question whether or not the classical capacity verifies any of the three properties of convexity, additivity, and computability. 

In some scenarios, sender and receiver may share additional resources which they can leverage to increase their communication rates. The capacities of a channel for a communication task assisted by additional resources turn out to have completely different properties than their unassisted counterparts.  
One such example is shared entanglement. The entanglement-assisted classical capacity of a quantum channel $\cent(\N)$ is both convex and additive and can be computed efficiently~\cite{Bennett_02}. 

Alternatively, one may consider 
the environment of the channel as a friendly helper that `assists' the sender during information transmission \cite{Winter_05}. This third party can input states independently of the sender or even interact with the sender by exchanging messages. This gives rise to a host of environment-assisted classical capacities depending on whether we have active or passive environment assistance \cite{Karumanchi_14} or whether the sender and environment are allowed to share entanglement or interact by means of local operations and classical communication. In our work, we focus on the weakest variant of assistance for classical communication when the helper is in the product state with the sender \cite{karumanchi_classical_2016}. The corresponding capacity is given by
\begin{equation}
\ch(\N) = \lim_{n\rightarrow\infty}\frac{1}{n}\max_{\eta}\cc^{(1)}(\N^{\otimes n}_{\eta}),
\end{equation}
where $\N^{\otimes n}_{\eta}(\rho) = \tr_F W^{\otimes n}(\rho\otimes\eta)(W^{\otimes n})^\dagger$. $W^{AE\to BF}$ is an isometric extension of the channel such that: $\N^{A\rightarrow B}(\rho^A)=\tr_F W \rho^A\otimes \proj{0}^E W^\dagger$ and $\eta$ is a state of the system $E$ over $n$ uses of the channel.

\section{Private capacity} 

We first show that private capacity is non-convex. Let us first define two families of channels.
The first is the $d$-dimensional erasure channel $\E_{d,p}$. Its action is defined as follows:
\begin{equation}
\E_{d,p}(\rho)=(1-p)\rho + p \ketbra{e}{e}.
\end{equation}
That is, $\E_{d,p}$ takes the input to the output with probability $1-p$ and with probability $p$ it outputs an erasure flag. The private capacity of the erasure channel is known to be~\cite{Wilde_13}:
\begin{equation}
\cp\left(\E_{d,p}\right)=\max\{0,(1-2p)\log d\}.
\end{equation}

The second is the `rocket channel' $R_d$. It was introduced by Smith and Smolin in~\cite{Smith_09b}. It takes two $d$-dimensional inputs that we label $C$ and $D$. The channel chooses two unitaries $U$ and $V$ at random~\footnote{The unitaries $U$ and $V$ are chosen uniformly at random from a unitary 2-design. 
Choosing the Clifford group or any other finite unitary 2 design has the advantage that the output of the channel is finite dimensional.} and applies them to $C$ and $D$ respectively, followed by the application of a joint dephasing operation $P$. The map is given by $P=\sum_{ij}\omega^{ij}\ketbra{i}{i}\otimes\ketbra{j}{j}$ with $\omega$ being a primitive $d$-th root of unity. Finally, the first system is traced out and the second system together with a classical description of $U$ and $V$ is sent to the receiver. Given $U$ and $V$ the action of the channel can be written as 
\begin{equation}
R_d^{UV}(\rho)=\tr_{C}\left(PUV\rho^{CD} \left(PUV\right)^*\right)\otimes\ketbra{U}{U}\otimes\ketbra{V}{V}\ ,
\end{equation}
where $PUV=P\cdot (U\otimes V)$. 
The total action of the channel is the average
\begin{equation}
R_d(\rho)=\mathbb E_{UV}R_d^{UV}(\rho).
\end{equation}

Rocket channels have small classical capacity for $d\geq 9$~\cite{Smith_09b}:
\begin{equation}
0<\cc(R_d)\leq 2.
\end{equation}
Now let us consider a convex combination of a flagged erasure channel and a flagged rocket channel:
\begin{equation}
\N_{q,d,p} = q\N^1_{d,p}+ (1-q)\N^2_d.
\end{equation}
where $\N^1_{d,p}=\E_{d^2,p}\otimes \ketbra{0}{0}$ and $\N^2_d=R_d\otimes \ketbra{1}{1}$.

In the following we prove that for some ranges of $d$, $p$ and $q$
\begin{equation}
\label{eq:target}
\cp(\N_{q,d,p})>q\cp(\N^1_{d,p})+(1-q)\cp(\N^2_{d}).
\end{equation}
The right-hand side of \eqref{eq:target} is bounded from above by
\begin{equation}
\label{eq:converse}
q\cdot\max\{0,(1-2p)2\log d\} + 2(1-q).
\end{equation}
We can bound $\cp(\N_{q,d,p})$ from below by $\cq(\N_{q,d,p})$. Hence, we can argue that any achievable rate for quantum communication (itself a lower bound on the quantum capacity) is a lower bound on the private capacity. Let $\rho^{A^1A^2C_1D_1C_2D_2}$ be some input for two uses of channel $\N_{q,d,p}$. Then:
\begin{align}
\cp(\N_{q,d,p})\geq\cq(\N_{q,d,p})\geq\frac12 \ic\left(\N_{q,d,p}^{\otimes 2},\rho\right).
\end{align}
Now, let the input be:
\begin{equation}
\label{eq:input}
\rho^{A^1A^2C_1D_1C_2D_2} = \Phi^{A^1D_1}\otimes\Phi^{C_1C_2}\otimes\Phi^{A^2D_2},
\end{equation}
where $\Phi^{AB}$ represents a maximally entangled state between systems $A$ and $B$. We use a subscript if the register corresponds to a concrete channel use and a superscript to number the subsystem: $C_2^1$ stands for the first subsystem of the register $C$ in the second use of the channel and $A^2$ the second subsystem of an auxiliary register $A$.

\begin{figure}
\begin{center}\includegraphics[width=8cm]{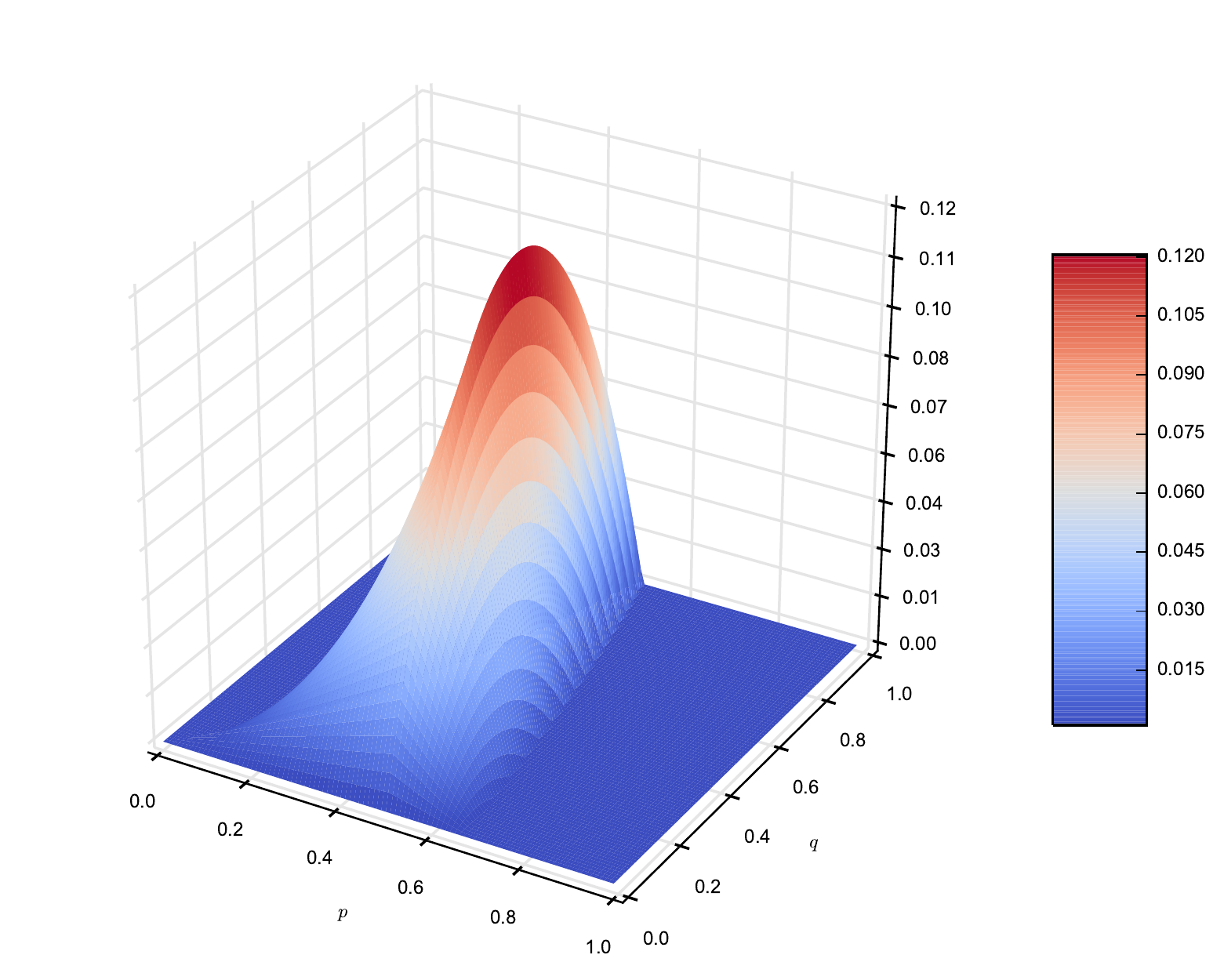}\end{center}
\caption{The figure shows the difference between \eqref{eq:achiev} and \eqref{eq:converse} normalized by $\log d$ when $d$ goes to infinity. A value larger than zero implies non-convexity of $\cp$. }
\label{fig:nonconvexity}
\end{figure}

The coherent information achieved by~\eqref{eq:input} is:
\begin{equation}
\ic(\N_{q,d,p}^{\otimes 2},\rho) = 2q((1-q)(2-3p)+q(1-2p))\log d.
\end{equation}
See the Supplemental material for details.
Consequently, the private capacity of $\N_{q,d,p}$ is bounded from below by:
\begin{align}
\cp(\N_{q,d,p})&\geq \frac{1}{2}\ic(\N_{q,d,p}^{\otimes 2},\rho) \\
                       &\geq q\left((1-q)(2-3p)+q(1-2p)\right)\log d.\label{eq:achiev}
\end{align}
It remains to compare the achievable bound in \eqref{eq:achiev} with the converse bound in \eqref{eq:converse}. For any triple $(q,d,p)$ such that \eqref{eq:achiev} is strictly greater than \eqref{eq:converse} the private capacity is non-convex. Figure~\ref{fig:nonconvexity} depicts the achievable region for which we exhibit non-convexity. 

\section{Classical environment-assisted capacity} 

We now turn to non-convexity of classical capacity with the weakest environment assistance. We start with providing two channels and a special entangled input state which we use to demonstrate this effect.
Consider a flagged combination of the two channels used in~\cite{karumanchi_classical_2016} to show superadditivity of $\ch$. 

The first channel is defined by a controlled unitary $V^{AE\to FB} = \sum_{x,z}|xz\rangle^F\langle xz|^A\otimes (W(x,z))^{E\to B}$ 
where $W(x,z)=X(x)Z(z)$, $X(x)|j\rangle = |(x+j)\mod d\rangle$, $Z(z)|j\rangle=\omega^{zj}|j\rangle$ and $\omega$ is again the primitive $d$-th root of unity.
 
The second channel is a $\text{SWAP}$ channel: $\text{SWAP}(|\phi\rangle^{A}\otimes|\psi\rangle^{E}) = |\psi\rangle^{B}\otimes|\phi\rangle^{F}$.
 
Thus, our channels will have the form $\N_1 = |0\rangle\langle0|\otimes V^{AE\to BF}$ and $\N_2 = |1\rangle\langle1|\otimes \text{SWAP}^{AE\to BF}$. Fix $|A|=|F|=d^2, |E|=d, |B|=d$. 
In the following we prove that for some range of $p$:
\begin{equation}
\label{eq:target2}
\ch\left(p\N_1 + (1-p)\N_2\right)>p\ch(\N_1)+(1-p)\ch(\N_2).
\end{equation}
It follows from~\cite{karumanchi_classical_2016} that $C_H(\N_1) = \log d$ and $\ch(\N_2) = 0$. Hence, the right-hand side of \eqref{eq:target2} is bounded from above by 
\begin{equation}
\label{eq:convcaphelp}
p\ch(\N_1)+(1-p)\ch(\N_2)\leq p\log d\ .
\end{equation}

In order to bound from below the left-hand side of \eqref{eq:target2}, consider two uses of the channel $\M = p\N_1 + (1-p)\N_2$. Let the state of the environment be the maximally entangled state between $E_1$ and $E_2$: $\Phi^{E_1E_2}$ 
and the input state to the channel: 
\begin{equation}
\rho^{XA_1A_2}=\frac{1}{d^2}\sum_{i,j=0}^{d-1}\ketbra{ij}{ij}^X\otimes\ketbra{ij}{ij}^{A_1}\otimes\ketbra{ij}{ij}^{A_2}
\end{equation} 
Then, 
\begin{align}
\ch(\M^{\otimes 2}) &\geq I(X:B_1B_2)_{\M^{\otimes 2}(\rho)},
\end{align}
and since $\M$ is flagged, we can also divide the mutual information into the sum of the mutual information associated with each channel action. Let us compute the corresponding output states:
\begin{align}
\N_1^{\otimes 2}(\rho)&=\frac{1}{d^2}\sum_{i,j=0}^{d-1}\ketbra{ij}{ij}^X\otimes Z(j)\otimes Z(j)(\Phi^{B_1B_2})\label{eq:n1n1}\\
\N_1\otimes \N_2(\rho)&=\frac{1}{d^2}\sum_{i,j=0}^{d-1}\ketbra{ij}{ij}^X\otimes \id\otimes W(i,j)\left( \Phi^{B_1B_2}\right) \\
\N_2^{\otimes 2}(\rho)&=\frac{1}{d^2}\sum_{i,j=0}^{d-1}\ketbra{ij}{ij}^X\otimes \Phi^{B_1B_2}
\end{align}
Note that $\N_2\otimes \N_1(\rho)$ is just $\N_1\otimes \N_2(\rho)$ with $B_1$ and $B_2$ swapped. 
The state obtained from the action of $\N_1^{\otimes 2}(\rho)$ follows from the observation that 
$W(x,z)\otimes W(x,z) \Phi=\id\otimes W(x,z)^T W(x,z) \Phi=Z(j)\otimes Z(j) \Phi$. 

It is easy to verify that $I(X;B_1B_2)$ vanishes when $\N_2\otimes \N_2$ is applied and takes the value $2\log d$ when either $\N_2\otimes \N_1$ or $\N_1\otimes \N_2$ is applied. In the case of $\N_1\otimes \N_1$ we can bound the mutual information by:
\begin{align}
I(X;B_1B_2)_\rho=\left\{\begin{aligned}\log d &\textrm{ if $d$ is odd}\\ \log\ d/2 &\textrm{ if $d$ is even}\end{aligned}\right.\label{eq:n1}
\end{align} 

Let us justify \eqref{eq:n1}. The input state is a clasical-quantum state of the form: $\sum_{ij}\ketbra{ij}{ij}^X\otimes\rho_{ij}^{A_1A_2}$.
We can write explicitly the input states as: 
\begin{equation}
\rho_{ij}^{A_1A_2}=\proj{ij}^{A_1}\otimes\proj{ij}^{A_2}\ .
\end{equation}
If we apply the channel to an input state, we can conclude from \eqref{eq:n1n1} that the output does only depend on $j$ and it simplifies to:
\begin{align}
\Phi_{j}^{B_1B_2} &:=W(i,j)\otimes W(i,j)\Phi \\
         &= \frac{1}{\sqrt{2}}Z(j)\otimes Z(j)\sum_{i=0}^{d-1}\ket{ii}\\
                               &=\sum_{i=0}^{d-1}\omega^{2ji}\ket{ii}\ .
\end{align}
Let $0\leq a,b\leq d-1$ and $a\neq b$, we can check the orthogonality between two output states:
\begin{align}
\braket{\Phi_a}{\Phi_b}&=\frac{1}{d}\sum_{i,j=0}^{d-1}\omega^{-2ai}\omega^{2bj}\braket{ii}{jj}\\
                                    &=\frac{1}{d}\sum_{j=0}^{d-1}(\omega^{2(b-a)})^j.\label{eq:blah}
\end{align}
\eqref{eq:blah} is a geometric series. Then, if $\omega^{2(b-a)}-1\neq 0$: 
\begin{equation}
\braket{\Phi_a}{\Phi_b}=\frac{(\omega^{2(b-a)})^d-1}{(\omega^{2(b-a)})-1}=0\ .
\end{equation}
That is, $\Phi_a$ and $\Phi_b$ are orthogonal except if $\omega^{2(b-a)}=1$ and then $\Phi_a=\Phi_b$. This is the case if $d$ divides $2(b-a)$ which can only occur for $2(b-a)=d$. Hence if $d$ is even there are $d/2$ orthogonal states and if $d$ is odd there are $d$ orthogonal states. We conclude that $I(X;B_1B_2)$ equals $\log d$ if $d$ is odd and $\log d/2$ if $d$ is even as claimed.

Adding all the contributions we obtain for odd $d$:
\begin{align}
\label{eq:achiev2}
\ch(\M)&\geq \frac{1}{2}\left(2p(1-p)2\log d + p^2 \log d\right)\nonumber\\
           &=\left(2p-\frac{3}{2}p^2\right)\log d.
\end{align}

Finally, comparing the achievable bound in \eqref{eq:achiev2} with the converse bound in \eqref{eq:convcaphelp} one observes that for odd $d>1$ and $0<p<2/3$ the classical capacity with passive environment-assisted capacity is non-convex.

\section{Discussion}

Computability, additivity, and convexity are three fundamental properties of capacity which allow to characterize the usefulness of a quantum channel for a concrete communication task. 

Here, we focused our attention on non-convexity. 
Prior to our work, non-convexity had only been proven for the quantum capacity. We exhibit non-convexity of communication tasks involving classical information via quantum channels. 
Hence, our results show that non-convexity is a generic feature of communications over quantum channels that is not merely restricted to the transmission of quantum information. 
Furthermore, non-convexity is not an effect which concerns only a zero-measure set of quantum channels: by perturbing the channels in our construction one finds that the result still holds. However, it remains open how typical is non-convexity (and non-additivity) if one chooses two channels at random.  

Both our non-convexity proofs and that of the quantum capacity build on top of non-additivity proofs. It is unclear if this is an artifact of the constructions or they hint to a deeper relation between both properties. 
Moreover, the non-convexity functional that we introduce here gives rise to a potential capacity analogous to the one induced by non-additivity.  
It is tempting to conjecture that the the two potential capacities, and more broadly, non-convexity and non-additivity, are closely related. Hence, a better understanding of this relation might shed some light into how much do the different capacities really gauge the usefulness of quantum channels for communication tasks. 

\

{\bf Acknowledgments:} 
We thank Kenneth Goodenough, Fr\'ed\'eric Grosshans, Jonas Helsen and Stephanie Wehner for useful discussions and feedback.
SS acknowledges the support of Sidney Sussex College and European Union under project QALGO (Grant Agreement No. 600700). DE has been partially supported by STW, the NWO Vidi grant ``Large quantum networks from small quantum devices" and by the project HyQuNet (Grant No. TEC2012-35673), funded by Ministerio de Econom\'ia y Competitividad (MINECO), Spain.


%

\appendix

\widetext

\begin{center} \large{\textbf{Supplemental Material}} \end{center}

\section{Justification of \eqrefF.}
Now we analyze the coherent information achieved by the input 
\begin{equation}
\label{eq:input}
\rho^{A^1A^2C_1D_1C_2D_2} = \Phi^{A^1D_1}\otimes\Phi^{C_1C_2}\otimes\Phi^{A^2D_2}.
\end{equation}

After sending $\rho$ through two copies of the channel $\N_{q,d,p}^{\otimes 2}$, the resulting state is:
\begin{align}
\N_{q,d,p}^{\otimes 2}(\rho)&=q^2(\E_{d^2,p}\otimes \E_{d^2,p})(\rho)\otimes\ketbra{0}{0}\otimes\ketbra{0}{0}\nonumber\\
                       &\quad+q(1-q)(\E_{d^2,p}\otimes R_d)(\rho)\otimes\ketbra{0}{0}\otimes\ketbra{1}{1}\nonumber\\
                       &\quad+(1-q)q(R_d\otimes \E_{d^2,p})(\rho)\otimes\ketbra{1}{1}\otimes\ketbra{0}{0}\nonumber\\
                       &\quad+(1-q)^2(R_d\otimes R_d)(\rho)\otimes\ketbra{1}{1}\otimes\ketbra{1}{1}.
\end{align}
Since the channel is a flagged combination of $\cal E$ and $R$, the coherent information is just the weighted sum of four terms 
\begin{align}
\label{eq:cohinfsum}
\ic(\N_{q,d,p}^{\otimes 2},\rho)&=q^2\ic(\E_{d^2,p}\otimes \E_{d^2,p},\rho)\nonumber\\
                                              &\quad+q(1-q)\ic(R_d\otimes \E_{d^2,p},\rho)\nonumber\\
                                              &\quad+q(1-q)\ic(\E_{d^2,p}\otimes R_d,\rho)\nonumber\\
                                              &\quad+(1-q)^2\ic(R_d\otimes R_d,\rho).
\end{align}
By symmetry of the input state, one has
\begin{equation}
\ic(\E_{d^2,p}\otimes R_d,\rho)=\ic(R_d\otimes \E_{d^2,p},\rho) .
\end{equation} 
Let us compute each of the three terms. First, we consider two erasure channels. The resulting state is
\begin{align}
\Big(\id^{A^1A^2}\otimes\E^{C_1D_1\rightarrow B^1_1B^2_1}_{d^2,p}\otimes \E^{C_2D_2\rightarrow B^1_2B^2_2}_{d^2,p}\Big)(\rho) &= (1-p)^2\Phi^{A^1B^1_1}\otimes\Phi^{B_1^2B_2^2}\otimes\Phi^{A^2B^1_2}\nonumber\\
                                                      &\quad +p(1-p)\Phi^{A^1B^1_1}\otimes\mms^{B_1^2}\otimes\mms^{A^2}\otimes\ketbra{e}{e}^{B^1_2B^2_2}\nonumber\\
                                                      &\quad +p(1-p)\Phi^{A^2B^2_2}\otimes\mms^{B_2^1}\otimes\mms^{A^1}\otimes\ketbra{e}{e}^{B^1_1B^2_1}\nonumber\\
                                                      &\quad +p^2\mms^{A^1A^2}\otimes\ketbra{e}{e}^{B^1_1B^2_1}\otimes\ketbra{e}{e}^{B^1_2B^2_2}\ ,
\end{align}
where 
$\mms$ stands for the maximally mixed state. 
The four states of this mixture can be differentiated by checking the erasure flag. This implies that the coherent information can also be divided into the sum of the coherent information of each term.
\begin{align}
\label{eq:1}
\ic(\E_{d^2,p}\otimes \E_{d^2,p},\rho) &= (1-p)^2 2\log d  + 2p(1-p) 0+ p^2(-2\log d)\\
                                                                             &=(1-2p)2\log d.
\end{align}
The resulting state in the case of one erasure channel and one rocket channel is
\begin{align}
\label{eq:15}
(R_d\otimes\E_{d^2,p})(\rho) &= (1-p)(R_d\otimes \id)(\rho)+ p (R_d\otimes\E_{d^2,1})(\rho)
\end{align}
which yields
\begin{align}
\label{eq:2}
\ic(R_d\otimes \E_{d^2,p},\rho) &= (2-3p)\log d.
\end{align}
Finally, the use of two rocket channels yields 
\begin{equation}
\label{eq:3}
\ic(R_d\otimes R_d,\rho) \geq 0.
\end{equation}
For justification of \eqref{eq:2} and \eqref{eq:3} see Appendix \ref{ap:b}.

We plug \eqref{eq:1}, \eqref{eq:2}, and \eqref{eq:3} back into~\eqref{eq:cohinfsum}
\begin{equation}
\ic(\N_{q,d,p}^{\otimes 2},\rho) = 2q((1-q)(2-3p)+q(1-2p))\log d
\end{equation}

\section{Justification of \eqref{eq:2} and \eqref{eq:3}.}
\label{ap:b}
The arguments follow from \cite{Smith_09b}. Let us analyze the action of one rocket channel and one erasure. This action can be decomposed into the action of the identity channel with probability $(1-p)$ and an erasure with probability $p$ as stated in \eqref{eq:15}. Let us compute the resulting state in both situations. For $\rho^{A^1A^2C_1D_1C_2D_2} = \Phi^{A^1D_1}\otimes\Phi^{C_1C_2}\otimes\Phi^{A^2D_2}$ we get:
\begin{align}
(R_d\otimes \id)(\rho) &= R_d^{C_1D_1\rightarrow B_1}\left(\Phi^{A^1D_1}\otimes\Phi^{C_1B_2^1}\right)\otimes\Phi^{A^2B^2_2}
\end{align}
The key idea here is that the register $C_1$ is maximally entangled with the register $B_2^1$, which is available to the receiver. Hence, the receiver can undo each unitary applied to $C_1$ by applying the inverse of the transpose of the corresponding unitary. More precisely, for each choice of $U$ and $V$ from the channel:
\begin{align}
\left((V^\dagger)^{B_1} \circ P^{B_1B_2^1}\circ ((U^T)^\dagger)^{B_2^2} \circ R_d^{UV}\right)\left(\Phi^{A^1D_1}\otimes\Phi^{C_1B_2^1}\right)=\Phi^{A^1D_1}\otimes\pi^{B_2^1}
\end{align}

In the case of rocket channel and erasure we obtain:
\begin{align}
(R_d\otimes &\E_{d^2,1})(\rho)= R_d^{C_1D_1\rightarrow B_1}\left(\Phi^{A^1D_1}\otimes\pi^{C_1}\right)\otimes\pi^{A^2}\otimes\proj{e}^{B^1_2B_2^2}
\end{align}
Let us denote by $\Phi^{AB}_U=\left(\id\otimes U\right) \Phi^{AB} \left(\id\otimes U^\dagger\right)$, then: $\Phi^{AB}_{U^T}=\left(U\otimes \id\right) \Phi^{AB} \left(U^\dagger\otimes \id\right)$. 
If we focus our attention on the action of the rocket channel for some concrete $U$ and $V$:
\begin{align}
R_d^{UV}\left(\Phi^{A^1D_1}\otimes\pi^{C_1}\right) \label{rdaction}
&= \tr_{C_1}\left(\sum_{ijkl}\omega^{ij-kl}\proj{ij}^{D_1C_1}(\Phi_V\otimes\pi)\proj{kl}^{D_1C_1}\right)\\
&= \sum_{ijl}\omega^{i(j-l)}\proj{j}^{D_1}\Phi_V\proj{l}^{D_1}\nonumber\\
&= \sum_{j}\proj{j}^{D_1}\Phi_V\proj{j}^{D_1}\nonumber\\
&= \sum_{j}\left(V^T\otimes\proj{j}^{D_1}\right)\Phi\left(\left(V^T\right)^\dagger\otimes\proj{j}^{D_1}\right)\nonumber\\
&= U^T\otimes\deph\left(\Phi\right)\nonumber
\end{align}
where $\deph$ denotes the completely dephasing channel in the computational basis. We can conclude that $\ic(R_d\otimes \id,\rho)=2\log d$, $\ic(R_d\otimes \E_{d^2,1},\rho)=-\log d$ and $\ic(R_d\otimes \E_{d^2,p},\rho)=(2-3p)\log d$.

Now, let us analyze the action of two rocket channels. From the data processing inequality for coherent information we have that $\ic(R_d\otimes R_d,\rho)\geq \ic(\deph\circ R_d\otimes \deph\circ R_d,\rho)$. Now we will show that,
\begin{align}
\left[\id^{A^1}\otimes\deph\circ R^{UV}_d\right]&\otimes\left[\id^{A^2}\otimes \deph\circ R_d^{WX}\right]\left(\Phi^{A^1D_1}\otimes\Phi^{C_1C_2}\otimes\Phi^{A^2D_2}\right)\\
&= \left[\left(V^T\right)^{A^1}\otimes\deph\right]\otimes\left[\left(X^T\right)^{A^2}\otimes\deph\right]\left(\Phi^{A^1D_1}\otimes\Phi^{A^2D_2}\right),
\end{align}
where the action of $R^{UV}_d$ and $R^{WX}_d$ is described in~\eqref{rdaction}. 
Then, since coherent information is invariant under the application of local unitaries:
\begin{align}
\ic(R_d^{UV}\otimes R_d^{WX},\rho) &\geq \ic(\deph\otimes\deph,\Phi^{A^1D_1}\otimes\Phi^{A^2D_2}) =0\ .
\end{align}
We can write explicitly the form of the output after acting on the input state with $R_d^{UV}\otimes R_d^{WX}$:
\begin{align}
&\sigma^{A^1A^2D_1D_2}=\\&=\sum_{\substack{ijkl\\abcd}}\omega^{ij-kl+ab-cd}\left(\id\otimes\proj{j}\Phi_V^{A^1D_1}\id\otimes\proj{l}\right)\otimes \left(\id\otimes\proj{b}\Phi_X^{A^2D_2}\id\otimes\proj{d}\right)\tr\left(\proj{i}\otimes\proj{a}\Phi_{U^TW}\proj{k}\otimes\proj{c}\right)\\
&=\sum_{\substack{ijl\\abd}}\omega^{i(j-l)+a(b-d)}\left(\id\otimes\proj{j}\Phi_V^{A^1D_1}\id\otimes\proj{l}\right)\otimes\left(\id\otimes\proj{b}\Phi_X^{A^2D_2}\id\otimes\proj{d}\right)\bra{ia}\Phi_{U^TW}\ket{ia}.
\end{align}
If we apply a dephasing channel at the output we obtain the following:
\begin{align}
\deph\otimes\deph\left(\sigma^{A^1A^2D_1D_2}\right)&=\sum_{xy}\id^{A^1A^2}\otimes\proj{xy}^{D_1D_2}\sigma^{A^1A^2D_1D_2}\id^{A^1A^2}\otimes\proj{xy}^{D_1D_2}\\
&=\sum_{iajb}\left(\id\otimes\proj{j}\Phi_V^{A^1D_1}\id\otimes\proj{j}\right)\otimes\left(\id\otimes\proj{b}\Phi_X^{A^2D_2}\id\otimes\proj{b}\right)\bra{ia}\Phi_{U^TW}\ket{ia}\\
&=\sum_{jb}\left(\id\otimes\proj{j}\Phi_V^{A^1D_1}\id\otimes\proj{j}\right)\otimes\left(\id\otimes\proj{b}\Phi_X^{A^2D_2}\id\otimes\proj{b}\right)\\
&=\left(V^T\right)^{A^1}\otimes\deph\otimes\left(X^T\right)^{A^2}\otimes\deph\left(\Phi^{A^1D_1}\otimes\Phi^{A^2D_2}\right).
\end{align}

\end{document}